\documentclass[11]{article}

\usepackage{amssymb,latexsym,amsmath}
\usepackage{amsmath}     
\usepackage{dsfont}
\usepackage{upgreek}
\usepackage{graphics}
\usepackage{graphicx}
\usepackage{lscape}
\usepackage{multirow}
\usepackage{amssymb}
\usepackage{dsfont}
\usepackage{hyperref}
\usepackage{authblk}
\usepackage[switch,columnwise]{lineno}
\usepackage{graphicx}
\usepackage{caption}
\usepackage{subcaption}
\usepackage{booktabs}
\usepackage{setspace}

\begin{document}
\title{An Alternative Discrete Skew Logistic Distribution}
\author[1*]{Deepesh Bhati}
\author[2]{Subrata Chakraborty}
\author[1]{Snober Gowhar Lateef}
\affil[1] {Department of Statistics, Central University of Rajasthan, }
\affil[2] {Department of Statistics, Dibrugarh University, Assam}
\affil[*] {Corresponding Author: deepesh.bhati@curaj.ac.in}
\maketitle

\begin{abstract}
In this paper, an alternative Discrete skew Logistic distribution is proposed, which is derived by using the general approach of discretizing a continuous distribution while retaining its survival function. The properties of the distribution are explored and it is compared to a discrete distribution defined on integers recently proposed in the literature. The estimation of its parameters are discussed, with particular focus on the maximum likelihood method and  the method of proportion, which is particularly suitable for such a discrete model. A Monte Carlo simulation study is carried out to assess the statistical properties of these inferential techniques. Application of the proposed model to a real life data is given as well.
\end{abstract}

\section{Introduction}
Azzalini A.(1985) and many researchers introduced different skew distributions like skew-Cauchy distribution(Arnold B.C and Beaver R.J(2000)), Skew-Logistic distribution (Wahed and Ali (2001)), Skew Student's t distribution(Jones M.C. et. al(2003)). Lane(2004) fitted the existing skew distributions to insurance claims data. Azzalini A.(1985), Wahed and Ali (2001) developed Skew Logistic distribution by taking $f(x)$ to be Logistic density function and $G(x)$ as its CDF of standard Logistic distribution, respectively and obtained the probability density function(pdf) as
\begin{equation*}
f(x,\lambda)=\frac{2e^{-x}}{(1+e^{-x})^{2}(1+e^{-\lambda x})};\quad -\infty<x<\infty,\quad -\infty<\lambda<\infty
\end{equation*}
They have numerically studied cdf, moments, median, mode and other properties of this distribution. Chakraborty et. al.(2002) following Huang et. al.(2007) considered and introduced a new skew Logistic distribution and studied its properties. They have also studied the CDF,the moments and presented data fitting examples. 

Researchers in many fields regularly encounter variables that are discrete in nature or in practice. In life testing experiments, for example, it is sometimes impossible or inconvenient to measure the life length of a device on a continuous scale. For example, in case of an on/off-switching device, the lifetime of the switch is a discrete random variable. In many practical situations, reliability data are measured in terms of the number of runs, cycles or shocks the device sustains before it fails. In survival analysis, we may record the number of days of survival for lung cancer patients since therapy, or the times from remission to relapse are also usually recorded in number of days. In this context, the Geometric and Negative Binomial distributions are known discrete alternatives for the Exponential and Gamma distributions, respectively. It is well known that these discrete distributions have monotonic hazard rate functions and thus they are unsuitable for some situations. 
On the other hand count data models such as Poisson, Geometric can only cater to positive integers along with zero values. But in some analysis often the interest lies not only in counts but in changes in counts from a given origin, in such situation the variable of interest can take either zero, positive or negative value. Again in many situations the interest may be in the difference of two discrete random variables each having integer support $[0,\infty)$. The resulting difference itself will be another discrete random variable but with integer support $(-\infty,\infty)$(see Chakraborty and Chakravarty(2016)).Such random variables arise in many situations in reliability theory, risk analysis, sports modelling etc. In all cases, a discrete random variable with integer support $(-\infty,\infty)$ is the most appropriate model to fit the data. One easy way to address this problem is to construct appropriate discrete by discretizing the underlying continuous distribution. There are different techniques available in the literature for the construction of appropriate discrete model from a continuous distribution (for a comprehensive review on this topic see Chakraborty(2015)).

Although much attention has been paid to deriving discrete models from positive continuous distributions, relatively less interest has been shown in discretizing continuous distributions defined on the whole set $R$, the few exceptions are the discrete Normal distribution introduced by Roy(2003),the discrete Laplace distribution by Kozubowski and Inusah(2006), Discrete Logistic distribution of Chakraborty and Charavarty(2016). The Skellam distribution(1946) which was derived as the difference between two independent Poisson is the earliest discrete model available for this type of data.

In this paper, a new discrete skew probability distribution with integer support $(-\infty,\infty)$ is proposed by discretizing the recently introduced two-parameter skew Logistic distribution $SLogistic(p,q)$ of Sastry and Bhati (2014). Important properties such as the cumulative distribution function, hazard function, quantile function, mean, median, variance, mode of the distribution are derived. The estimation of the parameters by using method of proportion and method of maximum likelihood is discussed. An algorithm for generating the skew Logistic random variable is presented along with some simulation studies. Parameter estimation by different methods and a data fitting is also studied.

\section{Proposed two parameter discrete skew Logistic distribution}
Here we first briefly introduce the skew logistic distribution recently introduced by Sastry and Bhati(2014) before proposing the discrete skew Logistic distribution.
\subsection{Continuous skew Logistic distribution}
The pdf of the $SLogistic(\kappa,\beta)$ with skew parameter $\kappa>0$ and scale parameter $\beta>0$ of Sastry and Bhati(2014) is given by
\begin{equation}
   f(x,\kappa,\beta) =  \frac{2\kappa}{1+\kappa^{2}}
   \begin{cases}
      \frac{e^{-\frac{x}{\kappa\beta}}}{\beta \left( 1+e^{-\frac{x}{\kappa\beta}}\right)^{2}} \quad & \text{if} \quad x < 0 \\
       \frac{e^{-\frac{x\kappa}{\beta}}}{\beta \left( 1+e^{-\frac{x\kappa}{\beta}}\right)^{2}} \quad & \text{if} \quad x\geq 0
     \end{cases}
\end{equation}
Letting $\kappa=1$, the model reduces to the standard(symmetric) logistic distribution(see Johnson et al. 2005), value of $\kappa<1$ leads to left-skewed logistic distribution whereas $\kappa>1$  leads to right-skewed logistic distribution. The corresponding cumulative distribution function (cdf) and survival function(sf) are respectively given as
\begin{equation}
   F(x,\kappa,\beta) =
     \begin{cases}
       \frac{2\kappa^{2}}{(1+\kappa^{2})}\frac{1}{1+e^{-\frac{x}{\kappa \beta}}}   \quad & \text{if} \quad x<0\\
\frac{\kappa^{2}}{1+\kappa^{2}}+\frac{2}{1+\kappa^{2}} \left( \frac{1}{1+e^{\frac{-x \kappa}{\beta}}}-\frac{1}{2}\right) \quad & \text{if} \quad x\geq 0
     \end{cases}
\end{equation}
and the survival function is
\begin{equation}
   S(x,\kappa,\beta) =
     \begin{cases}
1-\frac{2\kappa^{2}}{(1+\kappa^{2})}\frac{1}{1+e^{-\frac{x}{\kappa \beta}}}  \quad & \text{if} \quad x<0\\
\\
\frac{2}{1+\kappa^{2}} \left(\frac{e^{-\frac{x\kappa}{\beta}}}{1+e^{-\frac{x\kappa}{\beta}}}\right)  \quad & \text{if} \quad x\geq 0
     \end{cases}
\end{equation}

\subsection{New discrete skew Logistic distribution}
Roy (2003) first proposed the concept of discretization of a given continuous random variable. Given a continuous random variable $X$ with survival function (sf) $S_{X}(x)$, a discrete random variable $Y$ can be defined as equal to $\left\lfloor X \right\rfloor$ that is floor of $X$ that is largest integer less or equal to $X$. The probability mass function(pmf) $P[Y=y]$ of $Y$ is then given by
\begin{equation*}
P[Y=y]=S_{X} (y)-S_{X} (y+1)
\end{equation*}
The pmf of the random variable $Y$ thus defined may be viewed as discrete concentration (Roy(2003)) of the pdf of $X$. Such discrete distribution retains the same functional form of the sf as that of the continuous one. As a result, many reliability characteristics remain unchanged. Discretization of many well known distributions is studied using this approach (for detail see Chakraborty 2015). Notable among them are discrete normal distribution (Roy(2003)), discrete Rayleigh distribution(Roy( 2004)), discrete Maxwell distribution(Krishna and Pundir(2007)), discrete Burr (Krishna and Pundir, 2009, Khorashadizadeh et. al.(2013)) and discrete Pareto distributions (Krishna and Pundir, 2009), discrete inverse Weibull distribution (Jazi et al. (2009)), discrete extended Exponential(Roknabadi et al.(2009)), discrete Gamma distribution (Chakraborty and Chakravarty, 2012), discrete Log-Logistics (Khorashadizadeh et. al.(2013)), discrete generalized Gamma distribution (Chakraborty( 2015)) and discrete logistic distribution (Chakraborty and Chakravarty(2016)).

Using this concept, a two-parameter discrete probability distribution is proposed by discretizing the re-parametrized version of the two-parameter $SLogistic(\kappa,\beta)$ given in (1). First re-parameterization of Skew Logistic distribution in (1) is done by taking $p=e^{-\frac{\kappa}{\beta}}$  and $q=e^{-\frac{1}{\kappa\beta}}$, where $p$ and $q$ are related to $\kappa$ and $\beta$ and value of $\kappa$ and $\beta$ is \\
\begin{equation*}
\kappa=\sqrt{\frac{\log p}{\log q}} \quad and \quad \beta=\sqrt{\frac{1}{\log p \log q}}
\end{equation*} This lead us to the following definition of the proposed discrete logisctic distribution.

\noindent \textit{\textbf{Definition:}} A continuous random variable $X$ with survival function $S_{X} (x)$ is commonly said to follow the Discrete skew Logistic distribution with parameters $p$ and $q$ given as  $\mathcal DSLogistic(p,q)$ if its pmf $P[X=x]$ is given by

\begin{equation}
   P[X=x] =
     \begin{cases}
        \frac{2\log p}{\log(pq)}\left(  \frac{q^{-(x+1)}}{1+q^{-(x+1)}}-\frac{q^{-x}}{1+q^{-x}}\right)  \quad & \text{if} \quad  x=....,-2,-1 \\
         \frac{2 \log q}{\log(pq)}\left( \frac{p^{x}}{1+p^{x}}-\frac{p^{x+1}}{1+p^{x+1}}\right) \quad & \text{if} \quad  x=0,1,2,.....
     \end{cases}
\end{equation}\\
with corresponding cdf, reliability function and hazard function
\begin{equation}
   F_{X}(x) =
     \begin{cases}
        \frac{2 \log p}{\log(pq)}\frac{q^{-(x+1)}}{(1+q^{-(x+1)})} \quad & \text{if} \quad x=\cdots,-2,-1 \\
         1-\frac{2 \log q}{\log(pq)}\frac{p^{(x+1)}}{(1+p^{(x+1)})} \quad & \text{if} \quad  x=0,1,2,\cdots
     \end{cases}
\end{equation}
\begin{equation}
 S_{X}(x,p,q)= \begin{cases}
1-\frac{2 \log p}{\log(pq)}\frac{q^{-x}}{1+q^{-x}} \quad & \text{if} \quad   x= \cdots,-2,-1 \\
\frac{2\log q}{\log(pq)}\frac{p^{x}}{1+p^{x}} \quad & \text{if} \quad  x=0,1,2,\cdots
\end{cases}
\end{equation}
and
\begin{equation}
   h_{X}(x) =  \frac{p(x)}{S_X(x)}= 
     \begin{cases}
        \frac{2 \log p\left(1-q\right)}{(1+q^{x+1})\left(\log(pq)+q^{-x}\log(q/p)\right)} \quad & \text{if} \quad  x=\cdots,-2,-1 \\
         \frac{1-p}{1+p^{x+1}}   \quad & \text{if}  \quad  x=0,1,2,\cdots
     \end{cases}
\end{equation}\\
\textbf{Remark}: Letting $p=q$ in (4), the model reduces to a new probability distribution with integer support on $(-\infty,\infty)$(Chakraborty and Chakravarty(2012)), if $p>q$ it leads to right-skewed Logistic distribution whereas $p<q$ leads to left-skewed Logistic distribution.\\
Beside skew parameter $p$ and $q$, we can also introduce a location parameter $\mu$ thereby generalizing our proposed Discrete Skew Logistic model. Thus the resulting can be written as
\begin{equation}
   P[X=x|p,q,\mu] =
     \begin{cases}
        \frac{2 \log p}{\log(pq)}\left( \frac{q^{-(x-\mu+1)}}{1+q^{-(x-\mu+1)}}-\frac{q^{-(x-\mu)}}{1+q^{-(x-\mu)}}\right)  \quad & \text{if} \quad  x=....,-2+\mu,-1+\mu \\
         \frac{2 \log q}{\log(pq)}\left( \frac{p^{x-\mu}}{1+p^{x-\mu}}-\frac{p^{x-\mu+1}}{1+p^{x-\mu+1}}\right) \quad & \text{if} \quad x=\mu,\mu+1,\mu+2,.....
     \end{cases}
\end{equation}
Further it can be noted that the mode of the location family of Discrete Logistic distribution is either $\mu$ for $p >q$ and $\mu-1$ for $p<q$.\\
\begin{center}
\begin{figure}[h]
\includegraphics[scale=0.6]{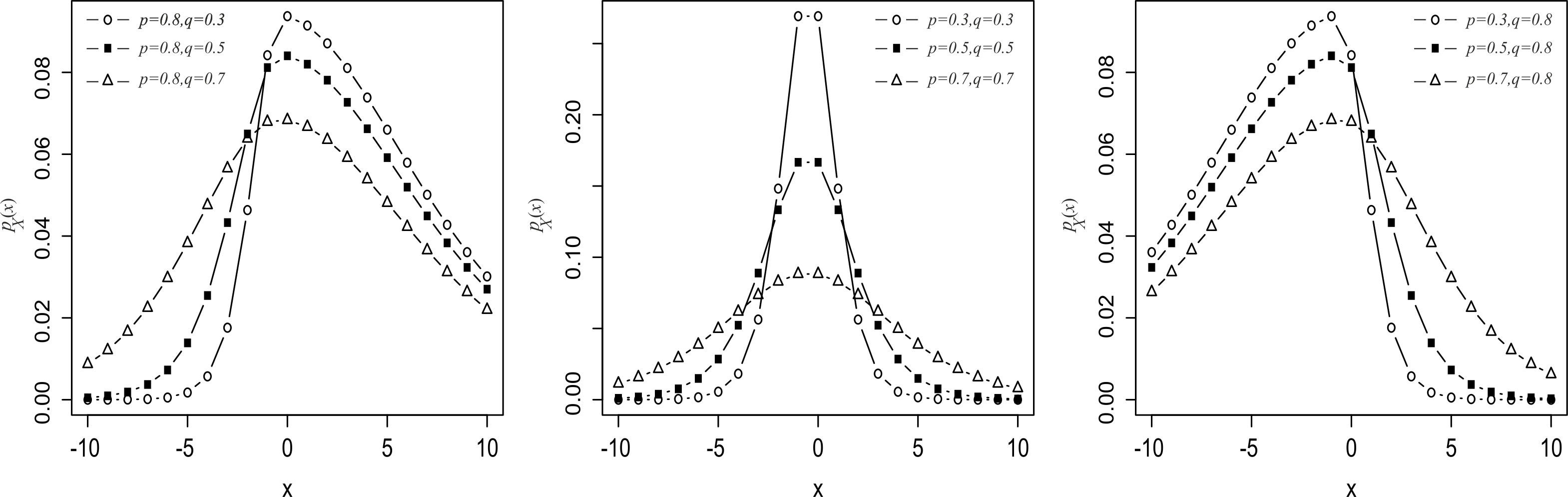}
\end{figure}
Figure 1: Probability mass function of $\mathcal DSLogistic(p,q)$ distribution for several combinations of the parameter values.
\end{center}

\section{Distributional properties}
\subsection{Quantile function}
The quantile of order $0<\gamma<1$, $x_\gamma$, can be obtained by inverting the cdf (5). Then if $\gamma \ge 1-\frac{2 \log q}{\log(pq)}\frac{p}{1+p}$,
\[
x_\gamma= \Big\lceil \log_p \left( \frac{(1-\gamma)\log(pq)}{2\log q-(1-\gamma)\log (pq)}\right)\Big\rceil-1
\]
\noindent where $\lceil z \rceil$ represents the smallest integer greater than or equal to $z$; otherwise for $\gamma < 1-\frac{2 \log q}{\log(pq)}\frac{p}{1+p}$
\[x_\gamma= \Big\lceil \log_q\left(\frac{2 \log p}{\gamma\log pq}-1\right)\Big\rceil-1\]

\noindent Further the median of $DSLogistic(p,q)$ obtained by substituting $\gamma=\frac{1}{2}$ in the above two equations
\begin{equation*}
Median=\begin{cases}
\Big\lceil \log_p \left(\dfrac{\log(pq)}{\log(\frac{q^{3}}{p})}\right)\Big\rceil-1 \qquad &\text{if} \quad q>p^{\frac{1+p}{3p-1}} \\
\Big\lceil \log_q \left(\dfrac{\log(\frac{p^{3}}{q})}{\log(pq)}\right)\Big\rceil-1\qquad &\text{if} \quad q\le p^{\frac{1+p}{3p-1}}
\end{cases}
\end{equation*}

\subsection{Moments}
Theorem 1: If $X \sim SLogistic(p,q)$, then \\
i) $E(Y)= \frac{2\log(\frac{p}{q})}{\log p \log q} \log_{e}(2)-0.05$ \\
ii)$V(Y)=\frac{(\log p)^3+(\log q)^3}{(\log pq) (\log p \log q)^2}\frac{\pi^2}{3}-\left(\frac{2\log(\frac{q}{p})}{\log p \log q} \log_{e}(2)\right)^{2}$ +0.0833 \\
\\
Proof: For continuous random variable following skew Logistic distribution with pdf in (1) it is known that
\begin{equation*}
E(X)=\frac{2\log(\frac{p}{q})}{\log p \log q} \log_{e}(2) \quad and \quad  V(X)= \frac{(\log p)^3+(\log q)^3}{(\log pq) (\log p \log q)^2}\frac{\pi^2}{3}-\left(\frac{2\log(\frac{q}{p})}{\log p \log q} \log_{e}(2)\right)^{2}
\end{equation*}
Now the discretized version $Y$ of $X$ that is $\mathcal DSLogistic(p,q)$ is defined as $Y=[X]$=largest integer less or equal to $X$ where $p=e^{-\frac{\kappa}{\beta}}$ and $q=e^{-\frac{1}{\kappa \beta}}$. Further, it can be assumed that $X=Y+U$, where $U$ is the fractional part of $X$ which is chopped off from $X$ to obtain $Y$. Now assuming $U$ is uniform in the support $(0,1)$ \\
\begin{equation*}
E(Y)=E(X)-E(U)
\end{equation*}
\begin{equation*}
E(Y)= \frac{2\log(\frac{q}{p})}{\log p \log q} log_{e}(2)-0.05 \quad since \quad E(U)=0.05
\end{equation*}
\begin{equation*}
V(Y)=V(X)+V(U)
\end{equation*}
\begin{equation*}
V(Y)=\frac{(\log p)^3+(\log q)^3}{(\log pq) (\log p \log q)^2}\frac{\pi^2}{3}-\left(\frac{2\log(\frac{q}{p})}{\log p \log q} \log_{e}(2)\right)^{2} +0.0833 \quad since \quad V(U)=0.0833\\
\end{equation*}
Hence proved.
\subsection{Monotonicity}
\begin{equation}
   \dfrac{f(x+1;p,q)}{f(x;p,q)}=
     \begin{cases}
          \frac{q^{-1}(1+q^{-x})}{(1+q^{-(x+2)})} \quad if \quad  x=....,-2,-1 \\
          \frac{p(1+p^{x})}{(1+p^{x+2})} \quad if \quad  x=0,1,2,.....
     \end{cases}
\end{equation}\\
Therefore, the above expression is monotone increasing for $x<0$ and monotone decreasing for $x\geq 0$.
\subsection{Mode}
\noindent \textbf{Theorem 2:} Discrete Skew Logistic distribution has unique mode at 0 if $p>q$, and at -1, if $p<q$ and two mode -1, 0 if $p=q$.\\
\textit{Proof:} Let us define $\Delta f_{X}(x)$ as
\begin{align*}
\Delta f_{X}(x)&=f(x+1)-f(x) \\
&=
\begin{cases}
\frac{2\log p}{\log(pq)}\left( \frac{q^{-x}}{1+q^{-x}}-\frac{2q^{-(x+1)}}{1+q^{-(x+1)}}+\frac{q^{-(x+2)}}{1+q^{-(x+2)}} \right)  \quad &\text{if} \quad  x=\cdots,-2,-1 \\
\frac{2\log q}{\log(pq)}\left(\frac{2p^{x+1}}{1+p^{x+1}}-\frac{p^{x+2}}{1+p^{x+2}}-\frac{p^{x}}{1+p^{x}} \right) \quad &\text{if} \quad x=0,1,2,\cdots 
\end{cases}
\end{align*}\\
It can be further observed that
\begin{equation*}
\Delta f_{X}(x)=
\begin{cases}
>0 \quad if \quad  x<0 \\
<0  \quad if \quad  x\geq 0
\end{cases}
\end{equation*}
Which implies that is monotonically increasing for $x<0$ and decreasing for $x\geq 0$. Moreover 0 is the unique mode if
\begin{equation*}
f(0)>f(-1)
\end{equation*}
\begin{equation*}
(\log q) \left( \frac{1+q}{1-q} \right)<(\log p) \left( \frac{1+p}{1-p} \right)
\end{equation*}
And -1 is the unique mode if
\begin{equation*}
f(-1)>f(0)
\end{equation*}
\begin{equation*}
(\log p) \left( \frac{1+p}{1-p} \right)<(\log q) \left( \frac{1+q}{1-q} \right)
\end{equation*}
Now, the function \[g(\xi)=\log \xi \left(\frac{1+\xi}{1-\xi} \right)\] is increasing when
$g'(\xi)>0$ i.e. $g'(\xi)=\frac{1-\xi^{2}+2\xi \log\xi}{\xi(1-\xi)^{2}}$\\
$g'(\xi)=(1-\xi^{2}) (1+\xi)+\xi\left[(1-\xi^{2})+\frac{(1-\xi^{2})^{2}}{2}+ \frac{(1-\xi^{2})^{3}}{3}+\cdots\right]$  is positive in $(0,1)$, since \\
$\xi\left((1-\xi^{2})+\frac{(1-\xi^{2})^{2}}{2}+ \frac{(1-\xi^{2})^{3}}{3}+\cdots \right)>0$. Thus it follows that $0$ is the unique mode if $p>q$,\\ -1 is the mode if $p<q$, and if $p=q$  there are two modes in 0 and in -1.

\section{Method of Estimation}
In this section we consider two method of estimation of parameters $p$ and $q$ namely (i) Method of proportion and zero's  (ii) Maximum likelihood method.
\subsection{Method of Proportion}
From (4), (5) and (6), we can obtain following probabilities as
\begin{equation}
p_0=P(X=0)= \frac{(1-p)\log q}{(1+p) \log pq}, \quad p^+=P(X\ge 0)= \frac{\log p}{\log pq} \quad \text{and} \quad p^-=P(X \le -1)= \frac{\log q}{\log pq}
\end{equation}
\noindent and solving these equations, we obtain
\begin{equation*}
{p}=\frac{p^+-p_0}{p^++p_0} \qquad \text{and} \qquad {q}=\left(\frac{p^+-p_0}{p^++p_0}\right)^{p^+/p^-}
\end{equation*}

\noindent Since a straightforward, estimate of $p_0$ is the proportion of sample values equal to zero to the total sample size, denote it with $r_0= \sum\limits_{i=1}^{n} \mathbb{I}_{\lbrace x=0\rbrace}/n$, analogously an estimate for $p^-$ and $p^+$ are the proportion of sample values less and greater than or equal to zero, i.e. $r^-= \sum\limits_{i=1}^{n} \mathbb{I}_{\lbrace x < 0 \rbrace}/n, r^+= \sum\limits_{i=1}^{n} \mathbb{I}_{\lbrace x \ge 0 \rbrace}/n$ respectively. Hence the estimates of $p$  and $q$ are
\begin{equation}
\tilde{p}=\frac{r^+-r_0}{r^++r_0} \qquad \text{and} \qquad \tilde{q}=\left(\frac{r^+-r_0}{r^++r_0}\right)^{r^+/r^-}
\end{equation}
As it is well know that the $r_0$, $r^-$ and $r^+$ are unbiased and consistent estimators of $p_0$, $p^{-1}$ and $p^+$ respectively, thus these can be used as an initial guess for searching global maxima of log-likelihood surface. Further this method fails to provide estimate of parameters if the sample contains no zero's or have all negative observations.
\subsection{Maximum Likelihood estimation}
The log-likelihood function of the discrete skew Logistic model computed on a iid sample $X_{1},X_{2},X_{3},\cdots,X_{n}$ is
\begin{eqnarray} \nonumber
l&=&n\log 2-n \log(\log pq)+ s^- \log(\log p)+s^+\log(\log q) \\ \nonumber
&&+\sum\limits_{i=1}^{n} \left( \log \left(  \frac{q^{-(x_i+1)}}{1+q^{-(x_i+1)}}-\frac{q^{-x_i}}{1+q^{-x_i}}\right) \cdotp \mathbb{I}_{\lbrace x_i<0\rbrace}\right) \\
&&+  \sum\limits_{i=1}^{n} \left(\log \left(  \frac{p^{x_i}}{1+p^{x_i}}-\frac{p^{x_i+1}}{1+p^{x_i+1}}\right) \cdotp \mathbb{I}_{\lbrace x_i \ge 0\rbrace}\right)
\end{eqnarray}
where $s^{-}=\sum_{i=1}^{n}I_{x_{i}<0}$ and $s^{+}=\sum_{i=1}^{n}I_{x_{i}\geq 0}$ expressing the number of negative and non-negative values in the sample.\\
Further differentiating the likelihood partially w.r.t $p$ and $q$, we get
\begin{eqnarray} \nonumber
\frac{\partial l}{\partial q}&=& -\frac{n}{q \log (p q)} +\frac{s^+}{q \log q} \\ \nonumber
&&+ \sum\limits_{i=1}^{n}\left(\frac{x_i+x_i q^{2 x_i+2}-2 q^{x_i+1}-(x_i+1)q^{2 x_i+1}-q (x_i+1)}{(1-q) q \left(q^{x_i}+1\right) \left(q^{x_i+1}+1\right)} \cdotp \mathbb{I}_{\lbrace x_i<0\rbrace}\right)  \\ \nonumber
\frac{\partial l}{\partial p}&=& -\frac{n}{p \log (p q)} +\frac{s^-}{p \log p} \\  \nonumber
&&+ \sum\limits_{i=1}^{n} \left(\frac{x_i+x_i p^{2 x_i+2}-2 p^{x_i+1}-(x_i+1) p^{2 x_i+1}-p(x_i+1)}{(1-p) p \left(p^{x_i}+1\right) \left(p^{x_i+1}+1\right)} \cdotp \mathbb{I}_{\lbrace x_i \ge 0\rbrace}\right)
\end{eqnarray}

The solution to above equations provides the maximum likelihood estimates(MLEs) of $p$ and $q$. It is quite clear that no close analytical expression can be derived for the MLEs, and have to be computed with some numerical procedure. We use the maxlik() function available in the R environment to carry out this task. \\
Further differentiating (12) and (13), we have
\begin{eqnarray*}
\frac{\partial^2 l}{\partial q^2}&=& n \left(\frac{1}{q^2 \log ^2(p q)}+\frac{1}{q^2 \log(p q)}\right)-s^+\left(\frac{1}{q^2 \log(q)^2}+\frac{1}{(q^2 \log(q)}\right)\\
&&+\sum\limits_{i=1}^{n}\left(\left(\frac{x_i}{q^2}+\frac{1-2 q}{q^2 (1-q)^2}+\frac{(x_i+1)^2}{q^2 \left(q^{x_i+1}+1\right)^2}-\frac{(x_i+1) (x_i+2)}{q^2 \left(q^{x_i+1}+1\right)} \right. \right. \\
&& \qquad \qquad  \left. \left. +\frac{x_i^2}{q^2 \left(q^{x_i}+1\right)^2}-\frac{(x_i+1) x_i}{qx_i^2 \left(q^{x_i}+1\right)}\right) \mathbb{I}_{\lbrace x_i<0\rbrace}\right)
\end{eqnarray*}
\begin{eqnarray*}
\frac{\partial^2 l}{\partial p^2}&=& n\left(\frac{1}{p^2 \log ^2(p q)}+\frac{1}{p^2 \log (p q)}\right) -s^- \left(\frac{1}{p^2 \log(p)^2}+ \frac{1}{p^2 \log(p)} \right)\\
&&+\sum\limits_{i=1}^{n}\left( \left( \frac{x_i}{p^2}+\frac{1-2 p}{p^2 (p-1)^2}+\frac{(x_i+1)^2}{p^2 \left(p^{x_i+1}+1\right)^2}-\frac{(x_i+1) (x_i+2)}{p^2 \left(p^{x_i+1}+1\right)} \right. \right. \\
&& \qquad \qquad \left. \left. +\frac{x_i^2}{p^2 \left(p^{x_i}+1\right)^2}-\frac{(x_i+1) x_i}{p^2 \left(p^{x_i}+1\right)}\right)\mathbb{I}_{\lbrace x_i \ge 0\rbrace} \right)
\end{eqnarray*}
Remembering that $E(I_{x\geq 0})=P[x\geq 0]=\frac{\log q}{\log(pq)}$ and $E(I_{x\leq 0})=P[x< 0]=\frac{\log p}{\log(pq)}$, we can compute the elements of $I(p,q)$ . If the true values of $p$  and $q$ are not available, one can plug in $\hat{p}_{ML}$ and $\hat{q}_{ML}$ for $p$ and $q$ respectively and the naïve large-sample confidence intervals at the nominal level $(1-\alpha)$  can be separately provided for $p$ and $q$  as $\hat{p}_{ML}\pm Z_{(1-\frac{\alpha}{2})}\sqrt{\hat{I}^{-1}_{(n)22}}$. Note that instead of using the inverse of the observed information matrix $\hat{I}^{-1}_{(n)11}$  one can use $\frac{I^{-1}}{n}$.

\section{Simulation study}
A Monte Carlo simulation is a broad class of computational algorithms that rely on repeated random sampling to obtain numerical methods. This study is conducted to assess the performance of the estimation method and asymptotic test. This method is useful for obtaining numerical solutions to problems that are complicated to solve analytically. A simulation study consisting of following steps is carried out for $(p,q)$ , where $p=0.25,0.50,0.75$ and $q=0.25,0.50,0.75$ and the values $n=25,50,75,100$ are the sample size. Under each scenario 1000 samples were drawn, the point estimates obtained with method of Maximum Likelihood(ML) and Method of Proportion(MP). MLEs of $\hat{p}$ and $\hat{q}$ are obtained by applying global numerical optimization method. It may be noted that the sample is obtained by generating Discrete Skew Logistic Distribution and then taking ceiling function. The estimation method has been compared through bias, mean square error(MSE) and the average width(aw)of 95\% of confidence limit of the obtained estimates over all $N$ samples. For these, the bias and mean square error(MSE) and the Proportion of $95\%$ confidence interval covering the true value of the parameter known as Coverage probability(CP), defined as

\begin{equation*}
bias(\hat{p})=\frac{1}{n}\sum_{i=1}^{n}(p_{i}-p_{0}), mse(\hat{p})=\frac{1}{n}\sum_{i=1}^{n}(p_{i}-p_{0})^{2}
\end{equation*}
and \\
\begin{equation*}
CP(\hat{p})=\frac{1}{n} \sum_{i=1}^{n}I(\hat{p}-1.96 SE <p_{0}<\hat{p}+1.96 SE)
\end{equation*}
where $SE$ is the standard error and is the true parameter. In the given table, it can be seen that the value of average bias, mean square error and the average width decreases with increase in the sample size $n$.
In this subsection, we assess the performance of ML estimators of  $\hat{p}$ and  $\hat{q}$ as sample size $n$, vary. For each of these sample sizes, we generate one thousand samples by using inversion method discussed above and obtain Maximum Likelihood estimators and standard errors of ML estimates, $(\hat{p},\hat{q})$ and $(s_{i, \hat{p}},s_{i, \hat{q}})$ for $i = 1, 2,\cdots, 1000$ respectively. For each repetition we compute bias, mean squared error, average width (aw) and Coverage Probability (CP). In the below table it can be clearly seen that the biasness, mean square error(MSE) and average width(aw) decreases with increase in and the confidence interval increases with increase in $n$.

\begin{landscape}
\begin{table}[]
\centering
\caption{Simulation study for biasness, mean square error(MSE) and average width(aw)}
\begin{tabular}{llllllllllllll} \hline
\multirow{2}{*}{Parameters} & \multicolumn{1}{c}{\multirow{2}{*}{n}} & \multicolumn{8}{c}{Method of Maximum Likelihood} & \multicolumn{4}{c}{Method of Proportion} \\ \cline{3-14}
 & \multicolumn{1}{c}{}  & bias(p) & bias(q) & mse(p) & mse(q) & aw(p)  & aw(q)  & CL(p)  & CL(q)  & bias(p)   & bias(q)  & mse(p)  & mse(q)  \\ \hline
\multicolumn{1}{c}{\multirow{4}{*}{p=0.25, q=0.25}} & 25  & -0.0077 & -0.0140 & 0.0057 & 0.0056 & 0.2836 & 0.2835 & 0.9029 & 0.9089 & 0.0189    & 0.0202   & 0.0139  & 0.0296  \\
\multicolumn{1}{c}{}       & 50   & -0.0063 & -0.0085 & 0.0029 & 0.0027 & 0.2050 & 0.2052 & 0.9320 & 0.9430 & 0.0046    & 0.0082   & 0.0061  & 0.0147  \\
\multicolumn{1}{c}{}       & 75   & -0.0032 & -0.0046 & 0.0018 & 0.0018 & 0.1690 & 0.1691 & 0.9500 & 0.9440 & 0.0041    & 0.0059   & 0.0039  & 0.0099  \\
\multicolumn{1}{c}{}       & 100  & -0.0035 & -0.0031 & 0.0014 & 0.0014 & 0.1467 & 0.1468 & 0.9440 & 0.9470 & 0.0040    & 0.0085   & 0.0032  & 0.0082  \\ \hline
\multirow{4}{*}{p=0.25,q=0.50} & 25  & -0.0169 & -0.0148 & 0.0081 & 0.0051 & 0.3222 & 0.2612 & 0.8980 & 0.9370 & 0.0136  & -0.0197  & 0.0229  & 0.0409  \\
                               & 50  & -0.0106 & -0.0074 & 0.0040 & 0.0022 & 0.2370 & 0.1849 & 0.9200 & 0.9500 & 0.0083  & 0.0022   & 0.0092  & 0.0181  \\
                               & 75  & -0.0062 & -0.0049 & 0.0026 & 0.0015 & 0.1959 & 0.1510 & 0.9330 & 0.9550 & 0.0017  & -0.0070  & 0.0063  & 0.0135  \\
                               & 100 & -0.0057 & -0.0022 & 0.0019 & 0.0012 & 0.1707 & 0.1305 & 0.9470 & 0.9400 & 0.0020  & 0.0011   & 0.0044  & 0.0094  \\ \hline
\multirow{4}{*}{p=0.25, q=0.75}& 25  & -0.0266 & -0.0072 & 0.0135 & 0.0017 & 0.4028 & 0.1526 & 0.8610 & 0.9496 & 0.0412  & -0.1115  & 0.0615  & 0.1015  \\
                              & 50   & -0.0197 & -0.0034 & 0.0076 & 0.0008 & 0.3055 & 0.1070 & 0.9010 & 0.9480 & 0.0224  & -0.0203  & 0.0250  & 0.0287 \\
                              & 75   & -0.0115 & -0.0031 & 0.0048 & 0.0005 & 0.2556 & 0.0874 & 0.9160 & 0.9450 & 0.0088  & -0.0100  & 0.0134  & 0.0117  \\
                              & 100  & -0.0088 & -0.0027 & 0.0038 & 0.0004 & 0.2227 & 0.0757 & 0.9280 & 0.9480 & 0.0091  & -0.0056  & 0.0100  & 0.0072 \\ \hline
\end{tabular}
\end{table}

\begin{table}[]
\centering
\caption{Simulation study for biasness, mean square error(MSE) and average width(aw) continue...}
\begin{tabular}{llllllllllllll} \hline
\multirow{2}{*}{Parameters} & \multicolumn{1}{c}{\multirow{2}{*}{n}} & \multicolumn{8}{c}{Method of Maximum Likelihood}    & \multicolumn{4}{c}{Method of Proportion} \\ \cline{3-14}
                                &   & bias(p) & bias(q) & mse(p) & mse(q) & aw(p)  & aw(q)  & CL(p)  & CL(q)  & bias(p)     & bias(q) & mse(p)  & mse(q) \\ \hline
\multirow{4}{*}{p=0.50,q=0.25}  & 25  & -0.0125 & -0.0198 & 0.0050 & 0.0079 & 0.2593 & 0.3239 & 0.9300 & 0.9000 & 0.054282331 & 0.0543  & 0.0723  & 0.0225 \\
                                & 50  & -0.0051 & -0.0106 & 0.0022 & 0.0039 & 0.1839 & 0.2375 & 0.9470 & 0.9340 & 0.041674184 & 0.0417  & 0.0484  & 0.0105 \\
                                & 75  & -0.0011 & -0.0037 & 0.0015 & 0.0027 & 0.1500 & 0.1965 & 0.9470 & 0.9300 & 0.042587804 & 0.0426  & 0.0480  & 0.0075 \\
                                & 100 & -0.0007 & -0.0052 & 0.0010 & 0.0018 & 0.1299 & 0.1709 & 0.9570 & 0.9460 & 0.044323072 & 0.0443  & 0.0506  & 0.0064 \\ \hline
\multirow{4}{*}{p=0.50, q=0.50} & 25  & -0.0066 & -0.0095 & 0.0061 & 0.0061 & 0.2934 & 0.2910 & 0.9310 & 0.9280 & 0.013149022 & 0.0131  & 0.0091  & 0.0263 \\
                                & 50  & -0.0031 & -0.0027 & 0.0027 & 0.0029 & 0.2053 & 0.2046 & 0.9470 & 0.9420 & 0.009081288 & 0.0091  & 0.0061  & 0.0122 \\
                                & 75  & -0.0008 & -0.0024 & 0.0019 & 0.0019 & 0.1675 & 0.1669 & 0.9470 & 0.9470 & 0.002198534 & 0.0022  & 0.0024  & 0.0071 \\
                                & 100 & -0.0013 & -0.0011 & 0.0013 & 0.0013 & 0.1447 & 0.1448 & 0.9550 & 0.9580 & 0.000993484 & 0.0010  & -0.0014 & 0.0056 \\ \hline
\multirow{4}{*}{p=0.50, q=0.75} & 25  & -0.0303 & -0.0112 & 0.0120 & 0.0020 & 0.3582 & 0.1631 & 0.9240 & 0.9520 & 0.032899581 & 0.0329  & -0.0072 & 0.0508 \\
                                & 50  & -0.0178 & -0.0050 & 0.0054 & 0.0009 & 0.2552 & 0.1132 & 0.9340 & 0.9520 & 0.007458131 & 0.0075  & -0.0034 & 0.0206 \\
                                & 75  & -0.0071 & -0.0042 & 0.0028 & 0.0006 & 0.2062 & 0.0925 & 0.9560 & 0.9520 & 0.004905407 & 0.0049  & -0.0054 & 0.0127 \\
                                & 100 & -0.0068 & -0.0018 & 0.0021 & 0.0004 & 0.1791 & 0.0794 & 0.9610 & 0.9370 & 0.002990232 & 0.0030  & -0.0040 & 0.0104 \\  \hline
\end{tabular}
\end{table}

\begin{table}[]
\centering
\caption{Simulation study for biasness, mean square error(MSE) and average width(aw) continue...}
\begin{tabular}{llllllllllllll} \hline
\multirow{2}{*}{Parameters} & \multicolumn{1}{c}{\multirow{2}{*}{n}} & \multicolumn{8}{c}{Method of Maximum Likelihood} & \multicolumn{4}{c}{Method of Proportion} \\ \cline{3-14}
                           &     & bias(p) & bias(q) & mse(p) & mse(q) & aw(p)  & aw(q)  & CL(p)  & CL(q)  & bias(p)   & bias(q)  & mse(p)  & mse(q)  \\ \hline
\multirow{4}{*}{p=0.75,q=0.25} & 25  & -0.0103 & -0.0295 & 0.0018 & 0.0135 & 0.1544 & 0.4008 & 0.9500 & 0.8500 & 0.0311 & 0.1050   & 0.0135  & 0.0815  \\
                           & 50  & -0.0057 & -0.0174 & 0.0008 & 0.0073 & 0.1079 & 0.3050 & 0.9460 & 0.9010 & 0.0286    & 0.0689   & 0.0071  & 0.0395  \\
                           & 75  & -0.0036 & -0.0087 & 0.0005 & 0.0051 & 0.0876 & 0.2545 & 0.9560 & 0.9170 & 0.0284    & 0.0636   & 0.0050  & 0.0291  \\
                           & 100 & -0.0016 & -0.0090 & 0.0004 & 0.0036 & 0.0753 & 0.2237 & 0.9350 & 0.9240 & 0.0332    & 0.0704   & 0.0043  & 0.0251  \\\hline
\multirow{4}{*}{p=0.75,q=0.50}           & 25  & -0.0118 & -0.0273 & 0.0020 & 0.0104 & 0.1633 & 0.3588 & 0.9550 & 0.9320 & 0.0149    & 0.0336   & 0.0162  & 0.0614  \\
                           & 50  & -0.0038 & -0.0158 & 0.0009 & 0.0050 & 0.1128 & 0.2547 & 0.9510 & 0.9370 & 0.0145    & 0.0246   & 0.0087  & 0.0330  \\
                           & 75  & -0.0032 & -0.0105 & 0.0005 & 0.0031 & 0.0921 & 0.2071 & 0.9460 & 0.9410 & 0.0062    & 0.0129   & 0.0053  & 0.0212  \\
                           & 100 & -0.0018 & -0.0070 & 0.0004 & 0.0022 & 0.0794 & 0.1790 & 0.9510 & 0.9470 & 0.0124    & 0.0226   & 0.0043  & 0.0167  \\ \hline
\multirow{4}{*}{p=0.75, q=0.75}  & 25  & -0.0068 & -0.0040 & 0.0025 & 0.0025 & 0.1845 & 0.1815 & 0.9360 & 0.9300 & 0.0141    & 0.0043   & 0.0229  & 0.0295  \\
                           & 50  & -0.0022 & -0.0029 & 0.0012 & 0.0011 & 0.1276 & 0.1282 & 0.9510 & 0.9520 & 0.0102    & 0.0039   & 0.0111  & 0.0146  \\
                           & 75  & -0.0018 & -0.0017 & 0.0007 & 0.0007 & 0.1042 & 0.1037 & 0.9370 & 0.9480 & 0.0014    & -0.0016  & 0.0074  & 0.0097  \\
                           & 100 & -0.0031 & -0.0011 & 0.0006 & 0.0005 & 0.0904 & 0.0898 & 0.9460 & 0.9520 & 0.0049    & 0.0009   & 0.0059  & 0.0078  \\ \hline
\end{tabular}
\end{table}
\end{landscape}

\section{Applications}
In this section, the $\mathcal DSLogistic$ distribution is applied to a data set given below (used by Kappenmann(1975) and Barbiero(2013) represents the difference between flood stage for two stations on the Fox River in Wisconsin
\begin{center}
1.96, 1.96, 3.60, 3.80, 4.79, 5.66, 5.76, 5.78, 6.27, 6.30, 6.76, 7.65,\\
7.84, 7.99, 8.51, 9.18, 10.13, 10.24, 10.25, 10.43, 11.45, 11.48, 11.75,\\
11.81, 12.34, 12.78, 13.06, 13.29, 13.98, 14.18, 14.40, 16.22, 17.06\\
\end{center}
We are assumes that this integer part of this random sample of size $n= 33$ are drawn from a Discrete skew Logistic distribution with a location parameter. To present a comparative performance appraisal here we consider three other discrete distribution namely, a new discrete Logistic distribution(Chakraborty and Chakravarty, 2016), discrete Laplace (Barbiero,2013), discrete Normal (Roy, 2003).\\
The pmfs of  new discrete Logistic of Chakraborty and Chakravarty, discrete Normal of Roy and discrete Laplace of Barbiero considered here for fitting are respectively given by\\
(i) New Discrete Logistic$(DLog(p,\mu))$\\
\begin{equation*}
p(Y=y)=\frac{(1-p)p^{y-\mu}}{(1+p^{y-\mu})(1+p^{y-\mu+1})}
\end{equation*}
where $y \in Z$, $-\infty< \mu <\infty$,$0<p<1$\\
(ii) Discrete Laplace$(DLaplace(p,q))$
\begin{equation*}
p(X=x)=\frac{1}{\log(pq)}
\begin{cases}
\log (p)(q^{-(x+1)}(1-q))  \quad & \text{if} \quad x<0\\
\log (q)(p^{x}(1-p)) \quad & \text{if} \quad x \geq 0
\end{cases}
\end{equation*}
where $0<p<1,0<q<1$\\
(iii) Discrete Normal of Roy$(DNR(\mu,\sigma))$\\
\begin{equation*}
P(Y=y)=\Phi\left( \frac{y+1-\mu}{\sigma}\right) -\Phi\left( \frac{y-\mu}{\sigma}\right)
\end{equation*}
\noindent $y=0,\pm 1,\pm 2,....\infty; -\infty< \mu <\infty, \sigma>0$, and  $\Phi(y)$ is the cdf of Standard Normal distribution.\\
Before employing them for our purposes, we transform them by subtracting their mode, $11.5$, and then taking the integer part. We expect that these final values can be modelled through our proposed discrete distribution. Maximum likelihood estimates of the parameters are obtained numerically by searching for global maxima of log likelihood surface using method of proportion estimates as initial values. \\
The findings are presented in Table() which reveals that all the distribution give satisfactory fits to this data set but the proposed $\mathcal DSLogistic(p,q)$ is the best fit in terms of log-likelihood values.

\begin{table}[]
\centering
\caption{My caption}
\label{my-label}
\begin{tabular}{crrrr} \hline
parameter(mse) & DSLog$(\mu,p,q)$ & DLog$(\mu,p)$ & DSLap$(p,q)$  & DNorm$(\mu,\sigma)$  \\ \hline
$\hat{\mu}$    & 11.5             & 9.382(0.753)  & -             & 9.318(0.717)         \\
$\hat{p}$      & 0.515(0.073)     & 0.664(0.039)  & 0.623(0.073)  &                      \\
$\hat{q}$      & 0.719(0.038)     &               & 0.791(0.036)  &                      \\
$\hat{\sigma}$ &                  &               &               & 4.11(0.507)          \\
$LogL$             & \textbf{-92.29}           & -94.62        & -94.38        & -93.57               \\   \hline
\end{tabular}
\end{table}

\section{Conclusions}
In this paper, a new discrete skew Logistic distribution defined on $\mathbb{Z}$ is proposed by discretizing a continuous Skew Logistic distribution. Some of its important probabilistic properties and parameter estimation is discussed. Monte Carlo simulation to investigae behaviour of the parameter estimation is also provided. From the results of the data fitting example considered here, the proposed discrete distribution is found suitable. Therefore, it may be conclude that the this new discrete distributions will be good competitor of the existing discrete distributions defined on $\mathbb{Z}$.

\end{document}